\documentclass{ws-ijmpa}

\begin{document}

%

\def\nocropmarks{\vskip5pt\phantom{cropmarks}}

\let\trimmarks\nocropmarks      

%


%
\catchline{}{}{}
%

\setcounter{page}{1}

\title{SPIN EFFECTS IN HIGH ENERGY FRAGMENTATION PROCESSES}

\author{LIANG Zuo-tang}

\address{Department of physics, Shandong University,  
Jinan, Shandong 250100, China}

\maketitle

\begin{abstract}
Recent measurements, in particular those on 
$\Lambda$ polarization and spin alignment of vector mesons in 
$e^+e^-$ annihilation at LEP, and those on the 
azimuthal asymmetry at HERA, have attracted
much attention on the spin effects 
in high energy fragmentation processes.
In this talk, we make a brief introduction 
to the different topics studied in this connection 
and a short summary of the available data. 
After that, we present a short summary of 
the main theoretical results that
we obtained in studying these different topics.
The talk was mainly 
based on the publications [5-9] which have 
been finished in collaboration with C.Boros, 
Liu Chun-xiu and Xu Qing-hua.
\end{abstract} 

\section{Introduction}

Spin effect is a powerful tool to study the properties of 
the hadronic interactions and hadron structures. 
Since the deep understanding of different aspects in  
spin physics of strong interaction almost always involves hadron 
production, spin effects in high energy fragmentation 
processes have attracted much attention recently, 
both experimentally and theoretically 
(see e.g.Refs.[1-14] and the references given there).
There are two main aspects in this connection, i.e.,
the dependence of the polarization and 
the momentum distribution of the 
produced hadron on the spin of the fragmenting quark. 
The former is usually referred as the spin transfer 
in the fragmentation process. 
For the latter, if the fragmenting quark is transversely 
polarized, we study the azimuthal angle dependence 
of the produced hadrons, 
and if the fragmenting quark is longitudinally polarized, 
we study a quantity which is called ``jet handedness''. 
The first two problems are closely related to the 
studies of the hyperon polarization in unpolarized 
hadron-hadron collisions and the left-right 
asymmetries in singly polarized hadron-hadron collisions\cite{LB2000}.
I will concentrate on these two problems in my talk. 

\section{Spin transfer in high energy fragmentation processes}

Spin transfer in high energy fragmentation process 
is defined as the probability for 
the polarization of the fragmenting quark to be transferred 
to the produced hadron. 
Here, we consider $q_0\to h(q_0...)+X$. 
We suppose that the $q_0$ was polarized 
before the fragmentation, and ask the following questions: 
(1) Will the $q_0$ keep its polarization?
(2) How is the relation between 
the polarization of $q_0$ and that of the 
produced $h$ which contains the $q_0$?
Clearly, the answers to these questions 
depend on the hadronization mechanism 
and on the spin structure of hadrons.
The study can provide useful information for both aspects. 
In particular, there exist now two distinctively different 
pictures for the spin contents of the baryons:
the static quark model 
using the SU(6) symmetric wave function  
[referred as the SU(6) picture],  
and the picture drawn from the data for polarized deeply inelastic 
lepton-nucleon scattering (DIS)
and SU(3) flavor symmetry in hyperon decay 
[referred as the DIS picture].  
It is natural to ask which picture 
is suitable to describe the question (2) mentioned above.
Obviously, the answers to these questions 
are also essential in the description of the 
puzzling hyperon transverse polarization 
observed already in the 1970s in unpolarized 
hadron-hadron reactions. 

\subsection{Hyperon polarization in high energy reactions as a tool to study the 
spin transfer in fragmentation processes}

It has been pointed out that\cite{BL98,LL2000} 
measurements of the longitudinal $\Lambda$ polarization 
in $e^+e^-$ annihilations at the $Z^0$ pole  
provide a very special check to the 
validity of the SU(6) picture in connecting 
the spin of the constituent to the polarization of 
the hadron produced in the fragmentation processes. 
This is because the $\Lambda$ polarization in this reaction 
obtained from the SU(6) picture should be 
the maximum among different models. 
Data are now available\cite{LamPol}
from both ALEPH and OPAL Collaborations. 
The results show that the SU(6) picture seems 
to agree better with the data\cite{LamPol} 
compared with the DIS picture. (See Fig.1).
This is rather surprising: the energy at LEP 
is very high hence the initial 
quarks and anti-quarks produced 
at the $e^+e^-$ annihilation vertices 
are certainly current quarks and current anti-quarks. 
They cannot be the constituent quarks used 
in describing the static properties of hadrons 
using SU(6) symmetric wave functions.
It is thus interesting and instructive to  
have further checks by making complementary measurements.  

\begin{figure}[htbp] 
\psfig{file=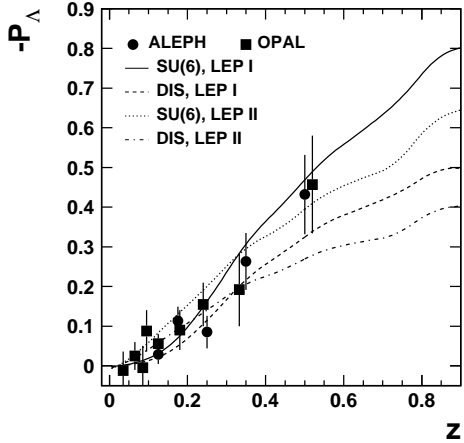,height=6.8cm}
\caption{Longitudinal $\Lambda$ polarization $P_\Lambda$ in 
$e^+e^-\to\Lambda X$ at LEP I and LEP II energies.}
\end{figure}

We note that, to study the spin transfer in 
fragmentation, we need to 
know the polarization of $q_0$ 
and measure the polarization of the produced $h$. 
Hence, hyperon productions in lepton-induced reactions 
are ideal to study this problem. 
Here, the polarization of quark can easily be calculated 
using the standard model for electroweak interaction 
and the hyperon polarization can easily be determined by measuring 
the angular distribution of its decay products.  
We have thus made a systematic study\cite{BL98}$^-$\cite{LXL2001} 
of hyperon polarizations in different lepton-induced reactions.
The obtained results can be used as further 
checks of the different pictures.  
Now we give a brief summary of the 
calculation method and the obtained results.

\subsubsection{The calculation mehod}

The calculation method has been formulated in different 
literature. Here, we summarize the main points in order 
to show the different inputs we need 
and what kinds of uncertaintities we 
may have in the calculations. 
 
We consider $q^0_f\to H_i+X$ and divide 
the produced $H_i$'s 
into the following groups:
(a) directly produced 
and contain the $q_f^0$'s; 
(b) decay products of heavier 
hyperons which were polarized before their decays; 
(c) directly produced but 
do not contain the $q_f^0$; 
(d) decay products of heavier hyperons 
which were unpolarized before their decays. 
Obviously, hyperons from (a) and (b) 
can be polarized while those from (c) and (d) are not. 
We obtain, 
\begin{equation}
P_{H_i}={ {\sum\limits_f t^F_{H_i,f} P_f \langle n^a_{H_i,f}\rangle
+\sum\limits_{j} t^D_{H_i, H_j} P_{H_j} \langle n^b_{H_i, H_j}\rangle}
 \over
{\langle n^a_{H_i}\rangle +\langle n^b_{H_i}\rangle + 
\langle n^c_{H_i}\rangle +\langle n^d_{H_i}\rangle} }. 
\end{equation}
The different quantities here are defined 
and obtained in the following way: 

(i) $P_f$ is the polarization of $q_f^0$ 
which is determined by the electroweak vertex.

(ii) $\langle n^a_{H_i,f}\rangle$ is the average number of 
$H_i$'s which are directly produced and contain 
$q_f^0$ of flavor $f$, and
$\langle n^b_{H_i,H_j}\rangle$ is that 
from the decay of $H_j$'s which were polarized;
$P_{H_j}$ is the polarization of $H_j$;
$\langle n^a_{H_i}\rangle$,
$\langle n^b_{H_i}\rangle$,
$\langle n^c_{H_i}\rangle$ and $\langle n^d_{H_i}\rangle$
are average numbers of $H_i$'s in group (a), (b), (c) 
and (d) respectively.
These average numbers of the hyperons of different 
origins are determined 
by the hadronization mechanisms and should be 
independent of the polarization of the initial quarks.
Hence, we can calculate them using a hadronization 
model which give a good description of the unpolarized data. 
We used Lund model implemented by JETSET or LEPTO in our calculations.

(iii) $t^F_{H_i,f}$ is the probability for 
the polarization of $q_f^0$ to be transferred 
to $H_i$ in group (a) and 
is called the polarization transfer factor, 
where the superscript $F$ stands for fragmentation. 
It equals to the fraction of 
spin carried by the $f$-flavor-quark 
divided by the average number of quark of flavor $f$ 
in $H_i$, which is different in the SU(6) 
or the DIS picture\cite{BL98,LL2000}.

(iv) $t^D_{H_i,H_j}$ is the probability for 
the polarization of $H_j$ to be transferred to 
$H_i$ in the decay process $H_j\to H_i+X$ and 
is called decay polarization transfer factor, 
where the superscript $D$ stands for decay. 
It is determined by the decay process and is independent 
of the process in which $H_j$ is produced.
For the octet hyperon decays, they are 
extracted from the materials in Review of Particle Properties. 
But for the decuplet hyperons, we have to use an estimation 
based on the static quark model. 
This is a major source of the theoretical uncertainties 
in our calculations of 
the final $P_{Hi}$'s in different reactions.

We applied\cite{BL98}$^-$\cite{XL2001} 
the method to $e^+e^-\to H_iX$, $\mu^-p\to \mu^-H_iX$ and 
$\nu_\mu p\to\mu^-H_iX$ at high energies and 
calculated the polarization of different octet hyperons 
in these reactions. 
We now summarize the main results as follows.

\subsubsection{The results for $e^+e^-$ annihilation}

For $e^+e^-\to H_iX$, 
we made the calculations at LEP I and LEP II energies. 
The results show that, all the octet hyperons should be 
significantly polarized 
and the polarizations are different in the SU(6) or the DIS picture. 
We also tried to make flavor separation. 
We found that it is impossible to separate 
only contribution from $u$ or $d$ to $\Lambda$. 
But we can enhance the contribution from $s$ fragmentation 
by giving some criteria to the selected events. 
For details, see Ref.[6].

\subsubsection{The results for deeply inelastic scattering at high energies}

In deeply inelastic lepton-nucleon scatterings, 
at sufficiently high $Q^2$ and hadronic energy $W$, 
hadrons in the current fragmentation region 
can be considered as the pure results of 
the fragmentation of the struck quarks.
There are two advantages to study hyperon polarization 
in $\mu^-p\to\mu^-H_iX$: 
Here, flavor separation 
can be achieved by selecting events in certain kinematic regions; 
and we can study the spin transfer both in longitudinally 
and in transversely polarized cases. 
We made the calculations for different combinations 
of beam and target polarizations. 
The results show the following characteristics:

(A) hyperons are polarized quite significantly 
if the beam is polarized but $\Lambda$ polarization 
is quite small in the case of unpolarized beam and polarized target. 

(B) there is significant contribution 
from heavier hyperon decay to $\Lambda$,  
it is even higher than the 
directly produced in most kinematic regions. 

(C) for $\Sigma^+$, the decay contribution is very small 
and the polarization is higher than that for $\Lambda$ 
and the differences from the different pictures are also larger. 

(D) the transverse polarization of the outgoing struck quark 
is obtained only in the case of using transversely polarized target.
But the resulted $\Lambda$ transverse polarization is very small 
and the decay influence is large. 
In contrast, the $\Sigma^+$ polarization 
is larger and there is almost no decay contribution. 

We made similar calculations for $\nu_\mu p\to\mu^-H_iX$. 
We found that there is a complete flavor separation for 
$\Sigma^+$ production. 
It comes almost completely from $u$ quark fragmentation.  
This leads to a quite high $\Sigma^+$ polarization. 
(See Fig.2.) 
However, for $\Lambda$ production, 
contribution from charmed baryon decay is very significant. 
It can completely destroy even the qualitative feature 
of the $\Lambda$ polarization. 
We thus reached the conclusion that, 
in deeply inelastic lepton-nucleon scattering,  
$\Sigma^+$ production is much more suitable to study 
the spin transfer in fragmentation processes. 
For details, see Ref.[7].

\begin{figure}[htbp] 
\psfig{file=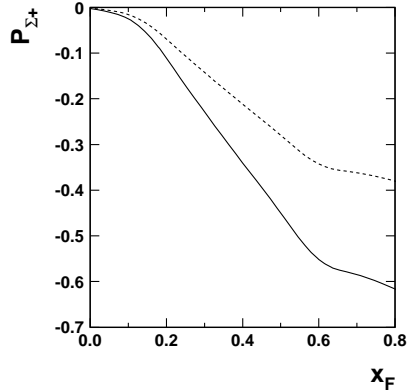,height=5cm}
\caption{$\Sigma^+$ polarization in  $\nu_\mu p$
$\to\mu^-\Sigma^+ X$ at high energies.
Here the solid and dashed lines 
are respectively the results obtained 
using the SU(6) and the DIS pictures.}
\end{figure}

\subsubsection{Hyperon plarization in $\nu_\mu\to\mu^-HX$ 
at the NOMAD energies}

We note that there are also measurements\cite{NOMAD} 
on $\Lambda$ polarization in $\nu_\mu N\to \mu^-\Lambda X$ by 
NOMAD collaboration at CERN. 
Compared the NOMAD data (See Fig.3) 
with the above-mentioned 
theoretical results, we see a distinct difference: 
While the theoretical results go to zero 
when $x_F$ goes to zero, the data show that 
$|P_\Lambda|$ rises monotonically when $x_F$ 
decreases from positive $x_F$ to negative $x_F$.
It does not go to zero when $x_F$ goes to zero. 
Does this imply that none of the pictures 
discussed above is suitable for $\nu_\mu N\to \mu^-\Lambda X$? 

A more detailed analysis shows 
that the answer to the question should be ``No!''
This is because, in the above mentioned calculations, 
we took only the struck quark fragmentation into account 
and neglected the influence from the remnant of the scattered nucleon. 
This is a good approximation only at high energies, 
or more precisely, at high $Q^2$ and $W$.
But, in the NOMAD experiments, 
the incident energies of the $\nu_\mu$ is $10\sim 50$ GeV and 
$W$ is only of several GeV. 
In this case, no separation between the fragmentation products of the 
struck quark and those of the nucleon remnant is possible. 
In particular, in the region of $x_F$ around zero, 
the contributions from the nucleon remnant can be very important. 
We studied this problem numerically 
using the event generator LEPTO. 
The results show that the contributions 
from the nucleon remnant indeed play an 
important role even at $x_F\sim 0.5$. 
We have to take it into account in such energy regions.

Using a valence quark model 
to calculate the polarization of the nucleon remnant, 
we made a very rough estimation 
of $P_\Lambda$ in $\nu_\mu N\to \mu^-\Lambda X$ 
at the NOMAD energies by taking both the fragmentation 
of the struck quark and that of the nucleon remnant into account. 
A qualitative agreement with the data\cite{NOMAD} is obtained 
(See Fig.3 or Ref.[7] for details).
 
\begin{figure}[htbp] 
\psfig{file=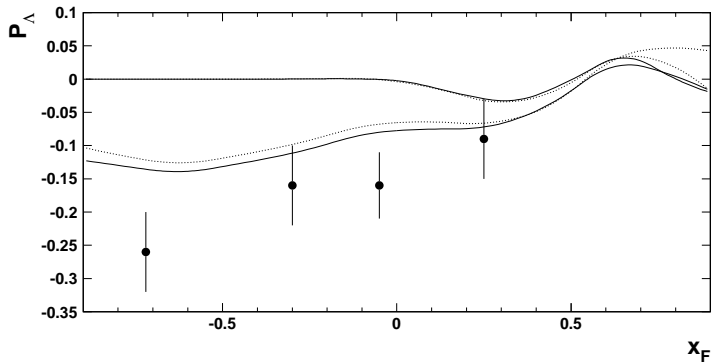,height=5cm}
\caption{$\Lambda$ polarization in $\nu_\mu N\to 
\mu^-\Lambda X$ at the NOMAD energies. 
Here, the upper and lower solid curves are respectively the 
results obtained using the SU(6) picture when the fragmentation of
the nucleon remnant is neglected or taken into account. 
Those dotted lines are the corresponding results 
using the DIS picture.}
\end{figure}

\subsubsection{Longitudinal hyperon polarization in 
high $p_\perp$ jets in polarized $pp$ collisions}

Last but not least, we emphasize that the
spin effects in fragmentation processes 
can also be studied in polarized $pp$ collisions 
(e.g. at RHIC) by measuring the hadrons 
in high $p_\perp$ jets. 
Here, it is envisaged that these hadrons are 
pure products of the fragmentation of the scattered 
quak (antiquark or gluon). 
Since we have many different hard subprocesses 
(e.g. $qq\to qq$, $qg\to qg$, or $gg\to gg$) 
which contribute to high $p_\perp$ jets, 
the situation here is much more complicated 
than that in the lepton induced reactions discussed above.
We have in particular the contribution from gluon fragmentation. 
This, on the one hand, make the study more difficult since
we know much less about gluon polarization and fragmentation. 
On the other, it makes the study also more interesting since 
we can use it to study not only quark but also gluon fragmentation. 

\nopagebreak

A simple Monte-Carlo study\cite{XLL2001} shows that, 
for moderately high $p_\perp$ (e.g. $3\sim 5$ GeV), 
gluon fragmentation dominates. This is a
kinematic region which is suitable for
studying gluon fragmentation. 
But, for very high $p_\perp$ and large $\eta$, 
quark fragmentation dominates.
Here, we can apply the above-mentioned
the method to calculate the longitudinal hyperon polarizations. 
Such calculations have also been carried out\cite{XLL2001}. 
An example of the obtained results is given in Fig.4. 

\begin{figure}[htbp] 
\psfig{file=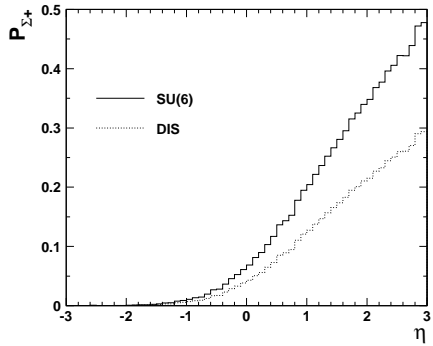,height=4.5cm}
\caption{$\Sigma^+$ polarization in high $p_\perp$
jets in $pp$ collisions at $\sqrt{s}=500$GeV
and $p_\perp >13$GeV.}
\end{figure}

\subsection{Spin alignment of vector mesons in high energy reactions}

Another aspect which is related to the 
problem of spin transfer in fragmentation is the 
spin alignment of vector meson in high energy reactions. 
It is clear that the polarization of the 
fragmenting quark can also be transferred to the vector meson $V$. 
This effect can be studied by measuring $\rho_{00}^V$, 
the $00$ component of the helicity density matrix of $V$, 
which is just the probability for $V$ to be in the helicity zero state. 
Data are available\cite{VecPol} for $e^+e^-$ 
annihilation at the $Z^0$-pole at LEP. 
It shows that,
for vector mesons with high momentum fraction, 
$\rho_{00}^V$ is much larger than $1/3$, 
the result expected in the unpolarized case. 

A simple calculation shows that\cite{XLL2001} 
these data\cite{VecPol} imply 
a significant polarization of the $\bar q$ that is
created in the fragmentation and combines 
with the polarized $q_f^0$ to form the vector meson. 
The polarization has a simple relation to that of the $q_f^0$, 
i.e., $P(\bar q)=-\alpha P(q_f^0)$,  
where $\alpha\approx 0.5$ is a constant. 
Using this we got a good fit to the data\cite{VecPol}. 
It should be interesting to see whether this relation 
is also true in other processes where polarized $q_f^0$ is produced. 
We thus apply it to other reactions and made predictions
for the spin alignments of vector mesons in these processes. 
They can be checked by future experiments. 
For details, see Ref.[8]. 

\nopagebreak
    
\section{Azimuthal asymmetry in the fragmentation of a transversely polarized quark}

In the fragmentation of a transversely polarized quark, 
the products can have an azimuthal asymmetry. 
The first measurement in this connection 
has been carried out by HERMES\cite{AziAsy},  
and the results show 
that such an effect indeed exists at that energy. 
What we would like to point out here is the following: 
The existence of such an azimuthal asymmetry 
is a direct consequence of Lund string fragmentation model due to 
conservation of energy-momentum and angular momentum.
The model was first used\cite{AGI79} to spin effects  
in 1979 to explain the unexpected 
$\Lambda$ polarization in unpolarzed pp collisions. 
By applying\cite{LB2000} it to the fragmentation of 
a transversely polarized quark, 
we obtain a significant azimuthal asymmetry. 
We are now working on the numerical calculations 
along this line in collaboration with the Lund group. 
Predictions for semi-inclusive deeply inelastic lepton-nucleon 
scattering and hadrons in high $p_\perp$ jets in transversely 
polarized $pp$ collisions will be available soon. 

\section*{Acknowledgements}

It is a great pleasure for me to thank the organizer 
for inviting me to give the talk. 
This work was supported in part by the Natural National Science Foundation 
(NSFC) and the Education Ministry of China.

\end{document}